\documentclass[aps,twocolumn,pra,showkeys,10pt]{revtex4-1}
\usepackage{graphicx}
\usepackage{amssymb}
\usepackage{amsfonts}
\usepackage{amsmath}
\usepackage{amsbsy}
\usepackage{natbib}
\usepackage{comment}
\usepackage[colorlinks=true, urlcolor=blue, citecolor=blue,
linkcolor=blue]{hyperref}
\usepackage{color}
\usepackage[utf8]{inputenc}
\usepackage[T1]{fontenc}
\usepackage{amsthm}
\usepackage{scalerel}

\usepackage{lmodern}
\usepackage{graphicx}
\usepackage{mathtools}
\usepackage{xcolor}

\newcommand{\bcorr}{ }

\begin{document}

\title{Multipolar, Polarization Shaped High Harmonic Generation by  Intense Vector Beams}

\author{Jonas W\"{a}tzel and Jamal Berakdar}
\affiliation{Institute for Physics, Martin-Luther-University Halle-Wittenberg, 06099 Halle, Germany}

\begin{abstract}
High harmonic generation (HHG) is a manifestation of the strongly  nonlinear response of matter  to intense  laser fields and has,  as the basis for  coherent XUV sources  a variety of applications. Recently, HHG from atoms in a phase and polarization structured laser was demonstrated and interpreted based on the transverse electric field component of the driving pulse. Here we point out that as dictated by Maxwell equations, such fields have a longitudinal component which in general has a fundamental influence on  the  charge dynamics. For instance, its interplay with the transversal field component enables endowing the emitted radiation locally with circular polarization and a defined polarity. It is shown that the time-dependent Stokes parameters defining the polarization state of HHG can be tuned  by varying  the waist of the driving field which  in turn, changes the ratio between  the longitudinal and transverse  electric-field components of the driving laser.
In addition, employing a multipole expansion  of the produced harmonics exposes the specific multipolar character and the relation to the spatial structure of the driving field polarization states.
The scheme proposed here allows a full polarization control of the emitted harmonics by only one driving laser. A tighter focusing of the driving pulse renders possible  the emission of harmonics with both even and odd spatial symmetry. The underlying mechanism is due to the fundamental interplay between the transverse and longitudinal components of the laser's electromagnetic vector potential. The ratio between those components is controllable by just focusing the laser spot, pointing to an accessible tool for polarization and polarity control of the high harmonics.
\end{abstract}

\maketitle

\section{introduction}

High harmonic generation (HHG) due to  highly non-linear light-matter interaction  \cite{agostini2004physics} paved the way for new types of XUV sources and ultrafast  (attosecond) spectroscopy \cite{krausz2009attosecond, popmintchev2010attosecond}.
 In recent years, driving with phase (optical vortices)  or polarization structured (vector beams)  fields has attracted much attention
  \cite{toda2010dynamics, zurch2012strong, hernandez2013attosecond, gariepy2014creating, geneaux2016synthesis, hernandez2017extreme, watzel2017tunable, paufler2019high, rego2019generation},  as the driving field characteristics allow to modulate the properties of the generated harmonics.
A particularly interesting  driving field is the radially  polarized   vector beam (RVB). Such RVB  can be  tightly  focused  \cite{chao2005soft} and are   attractive  for a number of applications, including  ultrafast diffraction  \cite{miao2015beyond} or lithography \cite{wagner2010euv, tallents2010optical}.  RVB are in general inherently non-transverse  \cite{zhan2009cylindrical} and may have a strong longitudinal electric field component \cite{lin2013longitudinal} for tightly focused beam  (cf. supp. mat.). This property is reflected in
a  new form  of light-matter interaction \cite{watzel2019magnetoelectric}, and hence  features  in HHG akin to the non-transverse RVB are to be expected, a case not yet clarified.
Another key point is that  the longitudinal and transversal electric fields $E^{(z)}_{\rm RVB}$ and $E^{(\rho)}_{\rm RVB}$ oscillate with a phase difference of $\pi/2$. Hence, for a tightly focused RVB, one can find positions in the beam spot with prevalent (local) circular polarization when $E^{(\rho)}_{\rm RVB}$ and $E^{(z)}_{\rm RVB}$ are of  the same magnitude (cf. appendix), a fact pointing to a possible polarization shaping of the HHG in a target driven by RVB. Indeed, the results presented here for HHG in  RVB-driven atomic ensemble   confirm the fundamental importance  of  the interplay between the transverse and  longitudinal components of RVB, an effect  tunable by  the laser focusing  that changes   the ratio between the two component amplitudes.
By doing so  the circular polarization of HHG, and in fact the spatially dependent Stokes parameters can be tuned. Other  effective methods for  circularly polarized HHG \cite{hickstein2015non, chen2016tomographic} use for instance bichromatic elliptically polarized pump beams \cite{fleischer2014spin} or counter-rotating few-cycle laser fields \cite{huang2018polarization}. Distinctive features of our HH are their spatially  multipolar character in addition to their polarization states.
The   symmetries of our HHs are analyzed below using a vectorial multipole expansion.
Structures  in the spectrum appear due to  the transversal (even symmetry) and the longitudinal (odd symmetry) components of the driving field. Hence, the focused RVBs provide a frequency-dependent tool for generating odd and even (X)UV harmonics.

\section{Theoretical model}

The vector and the scalar potentials of the  harmonics at the detector position $\pmb{r}_{\rm d}$
which are produced by an elementary (atomic) emitter at the position $\pmb{r}_{\rm i}$  are inferred  from the  laser-driven charge ($\rho_i$) and current ($\textbf{j}_i$) density distributions as
\begin{equation}
\pmb{A}_{\rm i}(\pmb{r}_{\rm d},t)=\mu_0/(4\pi)\int{\rm d}\pmb{r}'\,\pmb{j}_{\rm i}(\pmb{r}',t_R)/|\pmb{r}_{\rm d}-\pmb{r}_{\rm i}-\pmb{r}'|
\end{equation}
and
\begin{equation}
\Phi_{\rm i}(\pmb{r}_{\rm d},t)=1/(4\pi\epsilon_0)\int{\rm d}\pmb{r}'\,\rho_{\rm i}(\pmb{r}',t_R)/|\pmb{r}_{\rm d}-\pmb{r}_{\rm i}-\pmb{r}'|,
\end{equation}
where  $t_R=t-|\pmb{r}_{\rm d}-\pmb{r}_{\rm i}-\pmb{r}'|/c$ is the retarded time, and  $r_{\rm i}$ is the axial distance to the incident vector-beam optical axis   (which sets the $z$-axis of the global coordinate system).
\begin{figure}[t!]
\includegraphics[width=0.95\columnwidth]{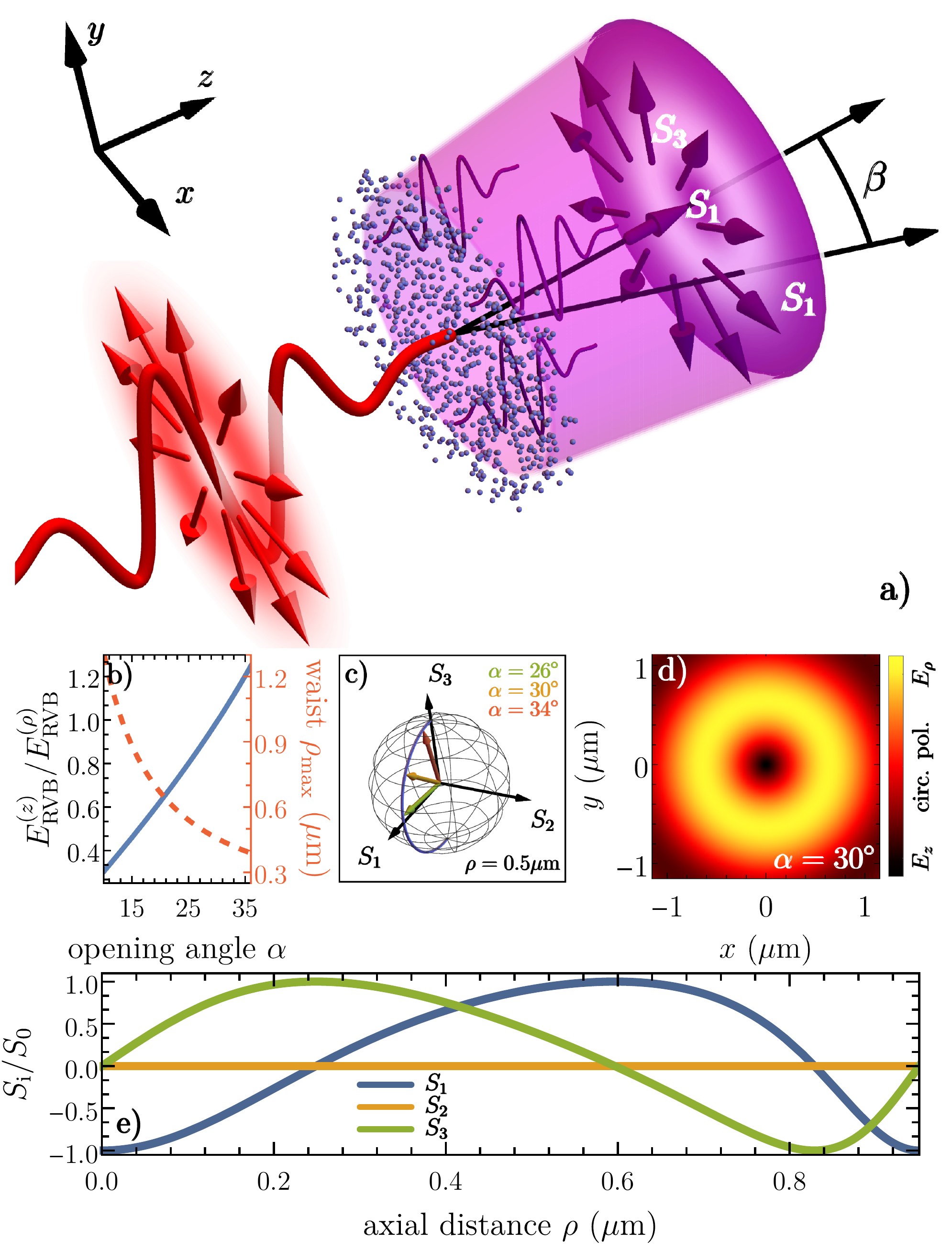}
\caption{High-harmonic generation driven by an intense  radially polarized laser. a) Schematic view of the HHG process:
an intense, radially polarized fs IR field is focused on a hydrogenic gas jet that responds with high harmonic emission.  The radiated field possesses   an inhomogeneous distribution of the Stokes parameters $S_{1,2,3}$. b)   (Blue curve)  incoming beam opening angle $\alpha$ dependence  of the  ratio  of its  on-axis  longitudinal component   $E^{(z)}_{\rm RVB}$ to   its    transverse component $E^{(\rho)}_{\rm RVB}$  (at $\rho_{\rm max}$).  Beam waist variation  is shown by the red curve . c)  Poincar\'{e} sphere for  different $\alpha$'s at an axial distance of $0.5\mu$m. d) color map of  polarization landscape in the focal plane $z=0$ for $\alpha=30^\circ$. e) Spatial dependencies of the normalized Stokes parameter $S_{\rm i}/S_0$ (${\rm i}=1,2,3$) corresponding to RVB with $\alpha=30^\circ$.}
\label{fig1}
\end{figure}
$\pmb{j}_{\rm i}(\pmb{r},t)$ and $\rho_{\rm i}(\pmb{r},t)$ of the individual atoms follow 
from a numerical propagation of the time-dependent three-dimensional Schr\"{o}dinger equation \cite{nurhuda1999numerical} involving the time-dependent Hamiltonian (we use atomic units for the quantum dynamics) 
\begin{equation}
\hat{H}_{\rm i}(t)=\left[\hat{\pmb{p}}-\pmb{A}_{\rm RVB}(\pmb{r}-\pmb{r}_{\rm i},t)\right]^2/2 + V(r_{\rm i}).
\end{equation}
Here, $\hat{\pmb{p}}$ is the momentum operator, and  we assume the target as a  gas of hydrogenic atoms, meaning  $V(r)=-1/r$ is the Coulomb potential. The vector potential of  RVB is taken as  a Bessel mode \cite{watzel2019magnetoelectric} so that a proper description of the electric longitudinal component is included automatically; using  Laguerre Gaussian modes (cf. Appendix) leads to the same conclusions drawn below. The total vector and scalar potentials  at the detector positioned at $\pmb{r}_{\rm d}$ is the  sum of the vector fields produced by the individual emitters:
\begin{equation}
\pmb{A}(\pmb{r}_{\rm d},t)=\sum_{\rm i}\pmb{A}_{\rm i}(\pmb{r}_{\rm d},t)~{\rm and}~\Phi(\pmb{r}_{\rm d},t)=\sum_{\rm i}\Phi_{\rm i}(\pmb{r}_{\rm d},t).
\end{equation}
The electromagnetic fields read  
\begin{equation}
\pmb{E}(\pmb{r}_{\rm d},t)=-\partial_t\pmb{A}(\pmb{r}_{\rm d},t)-\pmb{\nabla}_{\pmb{r}_{\rm d}}\Phi(\pmb{r}_{\rm d},t)
\end{equation}
and 
\begin{equation}
\pmb{B}(\pmb{r}_{\rm d},t)=\pmb{\nabla}_{\pmb{r}_{\rm d}}\times\pmb{A}(\pmb{r}_{\rm d},t).
\end{equation}
We confirmed, our scheme is equivalent to using Jefimenko's equations \cite{griffiths2018introduction, jefimenko1989electricity}. Phase-mismatch  effects may arise from dipole phase dependencies on the intensity, Gouy phase variation around the focal plane and dispersion effects in the neutral gas or in plasma. For optimal phase matching (phase mismatch of the $n$-th harmonic $k_n\rightarrow0$)  the gas jet is placed behind the RVB focal plane  \cite{hernandez2010high, garcia2013coherent}. Further, our HHG process is  carrier-envelope-phase insensitive \cite{hernandez2015carrier}. To focus on HHG, we assume a low-density target and suppress further discussions of optical refraction/propagation effects.

\section{Polarization control of HHG}

The opening angle $\alpha$ of the Bessel cone sets  the spatial extent in the focal plane (cf.  Fig.\,\ref{fig1}b) and the ratio between the peak longitudinal  (on the optical axis) and  transversal components (at the axial distance $\rho_{\rm max}$). Both components are important for the predicted effects. Increasing $\alpha$ tightens the spot size since (meaning $\rho_{\rm max}$
shrinks). The spatial inhomogeneity causes the Stokes parameters to become space-dependent. Due to   cylindrical symmetry, it is sufficient to investigate the Stokes parameters in the $x-z$ plane with the standard definitions \cite{mcmaster1954polarization}: $S_1$ and $S_2$ describe  linear polarization in the directions $\hat{e}_{x},\hat{e}_{z}$ and $(\hat{e}_{x}\pm\hat{e}_{z})/\sqrt{2}$ while $S_3$ signifies  circular polarization in the local plane.  Figure\,\ref{fig1}c) shows the  Poincar\'{e} sphere depending on  focusing (or on the opening angle $\alpha$) at an axial distance of $0.5\mu$m with astonishing implications. For a weak focusing ($\alpha=26^\circ$), we find that the polarization is nearly linear, characterized by $S_1\approx1$. Tightening the beam spot moves the Poincar\'{e} vector towards the poles indicating  \emph{circular} polarization. Furthermore, the vector always points on the meridian spanned by $S_1$ and $S_3$, signaling that linear polarization in the directions $(\hat{e}_{x}\pm\hat{e}_{z})/\sqrt{2}$ is suppressed. The polarization landscape in the focal plane is rather involved (cf. Fig.\,\ref{fig1}d):  Depending on the axial distance, we find a variation of the polarization state. While around the optical axis (where the longitudinal component dominates) the polarization points in the $z$-direction (Stokes parameter $S_1\approx-1$), we find a transition region where $E^{(\rho)}_{\rm RVB} \approx E^{(z)}_{\rm RVB}$.  The polarization state is circular because  both components oscillate with a phase difference of $\pi/2$.  Around  $\rho_{\rm max}$, the transversal component dominates, resulting in linear polarization perpendicular to the optical axis, meaning $S_1\approx+1$. Our strongest focusing  (at $\alpha=35^\circ$) corresponds to a FWHM of 1.2$\mu$m which is 1.5$\lambda$. Current focusing techniques are capable of generating vector beams with such tight focus  (even sub-diffraction focusing is possible) \cite{zhuang2019high}.\\
For illustrations  we run numerical simulations employing  a four-cycle long IR (800\,nm) radial vector beam with a  $\sin^2$ envelope and with  a     peak intensity at  $\rho_{\rm max}$  fixed  at $1.60\times10^{14}$\,W/cm$^2$,  independent of  opening angle $\alpha$.
\begin{figure}[t!]
\includegraphics[width=0.95\columnwidth]{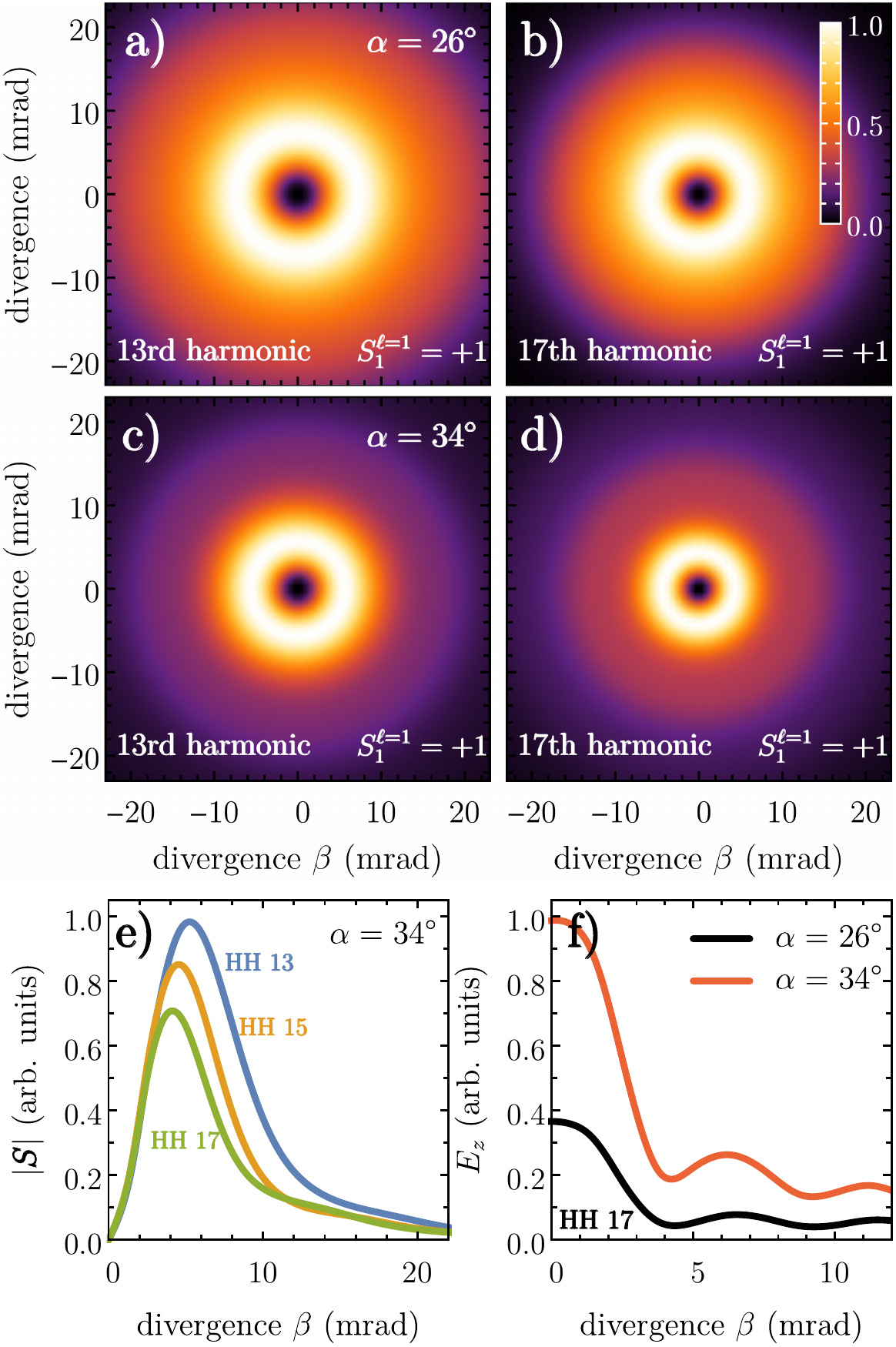}
\caption{Panels a-d): angular intensity profiles showing the absolute value of the Poynting vector of  two different harmonics (13 and 17) for two different laser focusing  (upper row $\alpha=26^\circ$; lower row $\alpha=34^\circ$). Panel e) demonstrates  the progress of divergency by increasing of the harmonic order. Panel f) shows the longitudinal field of the 17th harmonic for two different focusing setups of the incident RVB. The quantity $S^{\ell=1}_{1}=+1$ corresponds to the extended Stokes parameter for cylindrical beams and reveals pure radial polarization \cite{suzuki2015extended}.}
\label{fig2}
\end{figure}
The first four panels of Fig.\,\ref{fig2} present the angular profiles of the Poynting vector of two chosen harmonics for different  focusing of the incident beam. The far-field Poynting vector exhibits  a radial symmetry and a dark spot in the area around the optical axis, which can be explained by the diminishing   intensity  of the corresponding magnetic field when decreasing the axial distance. The polarization state is fully radial as inferred  from  calculating the extended (normalized) Stokes parameters $S_i^{\ell=1}$ for cylindrical beams \cite{suzuki2015extended}: Numerically, we found that $S^{\ell=1}_{1}=+1$ for all harmonics while the other two vanish.
The influence of tightening   the driving  RVB focus   is demonstrated  in  panels c-d). It can be concluded that the intensity of the outer area decreases for a larger opening angle $\alpha$. Figures\,\ref{fig2}e-f) present the diffraction properties of the HHG process: increasing the considered harmonic order, we find a tighter radiated beam spot since the axial distance to the intensity peak is lowered. However, the most exciting development  is due to the longitudinal component of the emitted electric far-field. As shown in panel f), a sharper focus leads to a strongly pronounced on-axis field with significant consequences for the polarization characteristics of the radiation.
Figure\,\ref{fig3}  reveals  the  polarization structure of the emitted radiation, evidencing that the relation between the longitudinal and transverse field components of the driving  field is transferred into the emitted  higher frequency fields. Due to cylindrical symmetry we confine the study of the polarization characteristics to the $x-z$ plane.
\begin{figure}[t!]
\includegraphics[width=0.97\columnwidth]{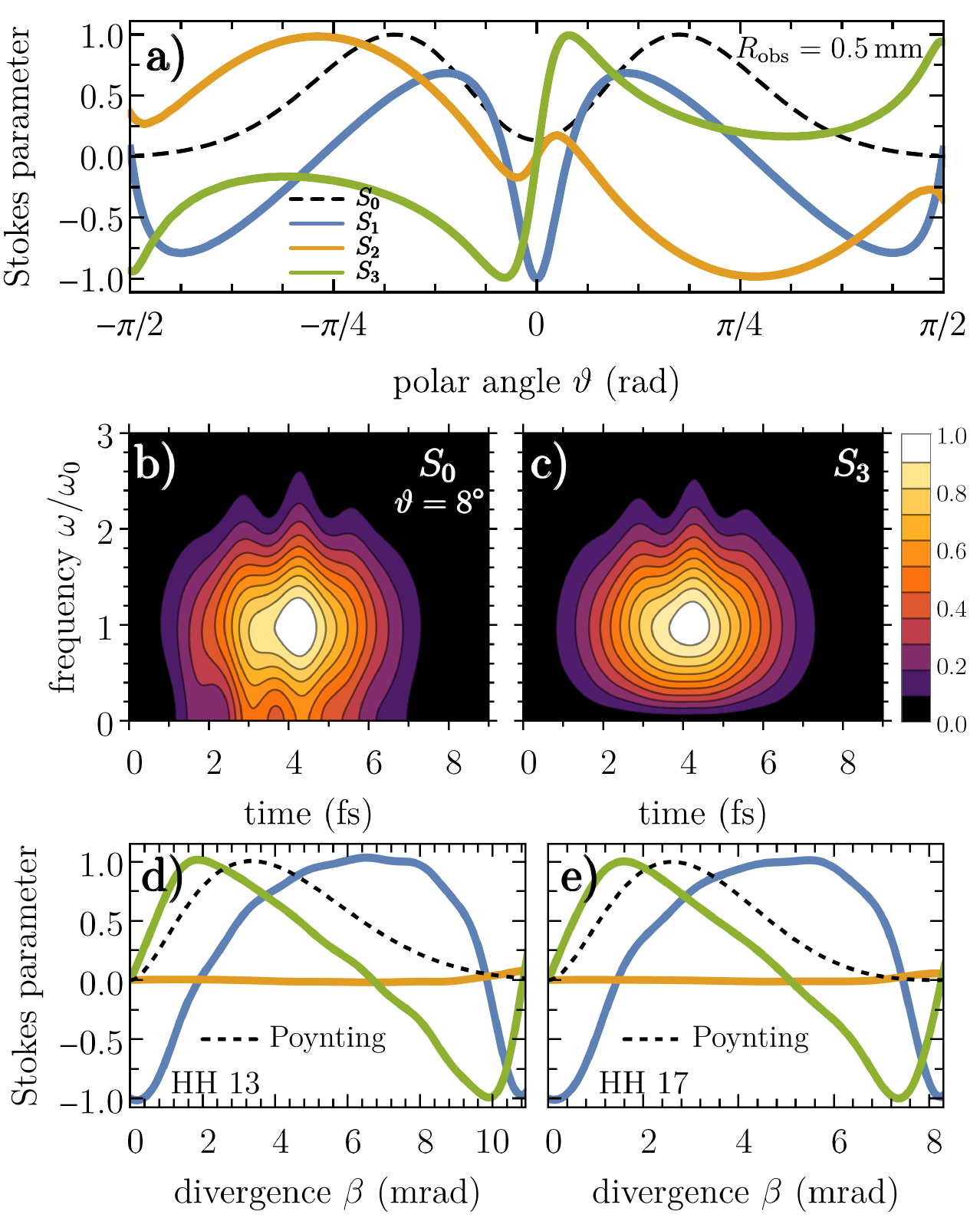}
\caption{Polarization landscape of  emitted radiation for an incident  RVB opening angle  $\alpha=30^\circ$. a) Angular-dependent Stokes parameters of the electric first harmonic order  at a sphere with radius $r_{\rm d}=0.5$\,mm. b-c) Time-dependent spectrum and third Stokes parameter of the electric far-field emitted in the asymptotic direction $\vartheta=8^\circ$, recorded at $r_{\rm d}=0.5$\,mm. d-e) (Normalized) Stokes parameters of the 13th and 17th harmonics as function of divergence angle.}
\label{fig3}
\end{figure}
At first, we introduce the polar angle $\vartheta$ as  the angle between the $z$-axis and the asymptotic direction of the detector position $\pmb{r}_{\rm d}$. In panel a), we present the four Stokes parameters, evaluated for the radiated $E_x(\pmb{r}_d,t)$ and $E_z(\pmb{r}_d,t)$ as a function of the polar angle for an observer distance $r_{\rm d}=0.5$\,mm. Noticeably, the polarization state changes continuously between linear (in $z-$,$\pm45^\circ$ and $x-$direction) and circular. Due to  symmetry, the radiation along the $z$-axis can only be $z$-polarized. A remarkable difference is the occurrence of the second Stokes parameter, which is absent in the incident RVB. Already at this stage, it is clear that (local) circular polarization can be observed in the far-field as a result of the coherent superposition of the emission of the individual radiators. At $\vartheta=\pm8^\circ$ we find highly distinctive circular polarization as evidenced  by (normalized) Stokes parameter $S_3=0.99$. The Stokes parameters nicely reflect the radial symmetry: The circular polarization changes its sign since under $E_x(-\vartheta)=-E_x(\vartheta)$ while $E_z(-\vartheta)=E_z(\vartheta)$ which corresponds to a phase jump of $\pi$ and reverses the direction. Hence, $S_3$ is an odd function. Note that $S_1+S_2+S_3=S_0$ meaning the emission is fully polarized.\\
The time-dependent buildup of the Stokes parameters \cite{eberly1977time, moskalenko2017charge} (via a wavelet Fourier transform) are presented in panel b-c). We choose the asymptotic direction of the radiation along  $\theta=8^\circ$ for an observer at  the distance  $r_d=0.5$\,mm  (maximal degree of circular polarization). The time-dependent spectrum of the electric far-field gives information about the duration of the emitted light pulse, which is in the fs regime. Furthermore, we find a symmetrical buildup and decay of the intensity ($S_0$) and the circular Stokes parameter ($S_3$). As already indicated in panel a), $S_0$ and $S_3$ are virtually equal in the whole time frame meaning that the radiation is always circularly polarized in this direction. We checked that $S_1$ and $S_2$ (not shown for brevity) are smaller then 0.05 at all times and frequencies, and hence, the (circular) polarization degree is persistently   $>0.99$. \\
A key finding is the possibility to endow the higher frequency regime with circular polarization, as shown in Figs.\,\ref{fig3}e-f) for the  spatially-dependent Stokes parameters of  the 13th and 17th harmonics for a varying divergence angle. In general, this behavior of $S_1, S_2$, and $S_3$ persists  for  higher harmonics in the (X)UV frequency regime. On-axis, the harmonics are strict linearly polarized, characterized by $S_1=-1$. The reason is the strong longitudinal component, which is discussed in Fig.\,\ref{fig2}. Increasing the axis distance results  in the  buildup of $S_3$ while $S_1$ decays. In reminiscence to the incident vector beam, as presented in Fig.\,\ref{fig1}e), $S_3$ decays again while $S_1$ changes its sign and approaches unity. Hence, the polarization state changes from linear (in the $z$-direction) to circular (relative to the $x-z$ plane) to linear (in the $x$-direction), meaning a transition into radial polarization when considering the whole beam spot. A  high degree of ellipticity is  around the maximum of the energy flux (Poynting vector), as indicated by the black, dashed curve.
\section{Multipolar HHG}

\begin{figure}[t!]
\includegraphics[width=0.97\columnwidth]{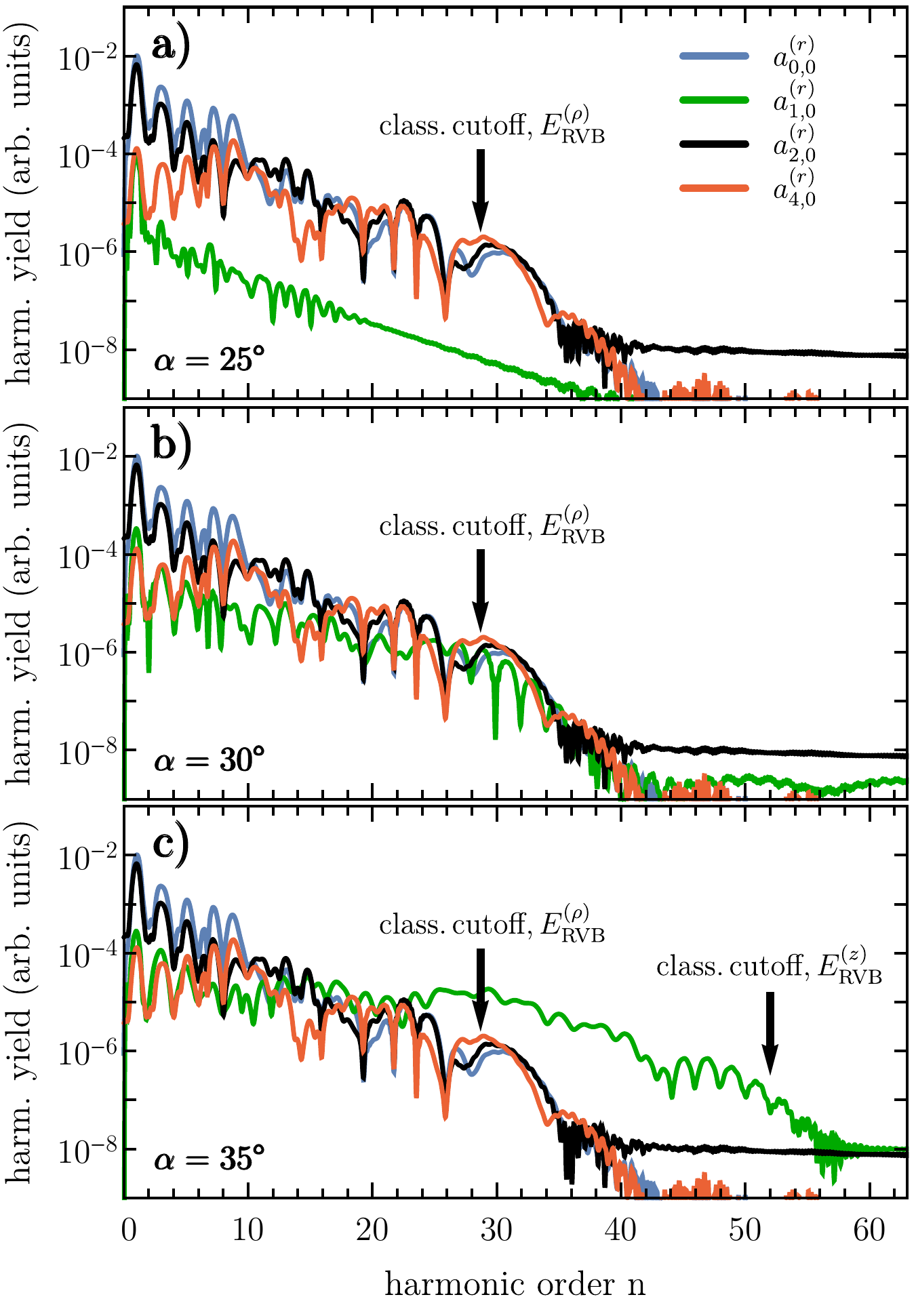}
\caption{Harmonic spectrum of the radiation produced by an incident RVB in dependence on the opening angle $\alpha$. The angular momentum dependent
multipole coefficients $a^{(r)}_{ L,0}$, as introduce in the text,   characterize the multipolarity of  the  emitted  harmonics of  order $n$.}
\label{fig4}
\end{figure}
For an insight into the polarity of the harmonics, we expand the electric far-field in vector spherical harmonics on a sphere with radius $r_{\rm obs}$: 
\begin{equation}
\begin{split}
\pmb{E}(\pmb{r}_{\rm obs},t)=&\sum_{L,M}a^{(r)}_{\rm L,M}(t)\pmb{Y}_{L,M} + a^{(1)}_{\rm L,M}(t)\pmb{\Psi}_{L,M} \\
& + a^{(2)}_{\rm L,M}(t)\pmb{\Phi}_{L,M}
\end{split}
\end{equation} 
(for conventions cf. Ref.[\onlinecite{barrera1985vector}]). All harmonics are independent of the azimuthal angle $\varphi$ due  to symmetry. Therefore, all coefficients with $M\neq0$ disappear (double-checked numerically). As indicated by  Fig.\,\ref{fig4}a), for a small opening angle $\alpha$, meaning a wide focusing, the transverse component $E_{\rm RVB}^{(\rho)}$ of the incident RVB is dominating the light-matter interaction with the result that the irradiated atomic layer emits radiation characterized by even multipoles. For $n > 10$ we find harmonic orders where the quadrupole (e.g., $n = 13$ and $15$) or even the hexadecapole (e.g. $n = 19$) are the leading multipole terms. The classical cutoff at $\hbar\omega_{\rm cutoff} = 3.17U_{\rm pond}$ (black arrow) is well reproduced as the harmonic yield decreases abruptly for $n>30$. \\
Note, atoms around the optical axis are exposed to $E_{\rm RVB}^{(z)}$, and the  dipole moment oscillates  in the $z$-direction. Thus, this part of the radiation   is dominated by the dipolar coefficients as for conventional  HHG. For the total HHG signal also   the transverse component is decisive which  oscillates with a $\pi/2$ phase difference. Panel (b)   evidences that a stronger focusing ($\alpha=30^\circ$) boosts the dipole coefficient $a^{(r)}_{1,0}$ drastically to a level close to the even coefficients.
A yet stronger focusing ($\alpha = 35^\circ$ in which case $E^{(z)}_{\rm RVB} > E^{(\rho)}_{\rm RVB}$)    has a substantial  impact on the harmonic spectrum (panel (c)):  Although the lower harmonics ($n< 10$) are still dominated by the even multipole coefficients $a^{(r)}_{1,0}$, from $n > 15$ the harmonics are strongly dipolar. Moreover, the whole HH cutoff is shifted by more than 20 orders, which can be explained by the larger classical cutoff corresponding to the incident longitudinal component $E^{(z)}_{\rm RVB}$.
Astonishingly, we can  produce (higher) harmonics with both parities by merely adjusting the waist of the driving pulse. The even multipole harmonics are a result of the radially polarized transverse  electric field component $E^{(\rho)}_{\rm RVB}$
revealing mirror reflection symmetry. In contrast, the linearly polarized $E^{(z)}_{\rm RVB}$ produces odd multipole harmonics, comparable to conventional atomic HHG. A third unusual attribute is the shift of the cutoff frequency by focusing, provided  the peak intensity of the RVB spot area is kept fixed.\\
{\bcorr HHG multipolarity  is relevant to spectroscopy as, depending on polarity, HHs induce transitions with different propensity rules. The electric field of the of a emitted harmonic of order $q$ can be expressed as
\begin{equation}
\pmb{E}_{q}(\pmb{r},t)=\sum_L a^{(r),q}_{L0}(r)\pmb{Y}_{LM}e^{-i\omega_{q}t}.
\end{equation}
A numerical analysis of the radiated vector potential $\pmb{A}(\pmb{r},t)$ reveals that it is approximately solenoidal. This is useful insofar as the coupling of $\pmb{A}(\pmb{r},t)$ to the charge current density of a finite size sample can be unitarily transformed to the following form ($\pmb{r}_j$ is the position of the electron):
\begin{equation}
\hat{H}_{\rm int}^{q}(t)=H_{\rm int}e^{-i\omega_{q}t}=\sum_{L,j}a_{L,0}^{(r),q}(r_j)\left(\pmb{r}_j\cdot\pmb{Y}_{L,0}\right)e^{-i\omega_{q}t}.
\label{eq_Hint_HH}
\end{equation}
For demonstration, let's consider an isotropic system amenable to an effective single-particle description. The ground state single-particle orbital can thus be written as  $\Psi_{i}(\pmb{r},t)=R_i(r)Y_{0,0}(\Omega_{\pmb{r}})$ with orbital energy $E_i$. A transition to an excited  state, presented by $$\Psi_{f}(\pmb{r},t)=R_f(r)Y_{\ell_f,0}(\Omega_{\pmb{r}})$$ with orbital energy $E_f$, is governed by the  matrix element
\begin{equation}
\begin{split}
\langle\Psi_f|H_{\rm int}|\Psi_i\rangle\propto&\frac{1}{\sqrt{4\pi}}\sum_{\ell_f}\sum_L \sqrt{(2\ell_f+1)(2L+1)}\\
&\times\mathcal{A}_{\ell_fL}
\begin{pmatrix}
\ell_f&L&0 \\ 0&0&0
\end{pmatrix}^2,
\end{split}
\end{equation}
where $$\mathcal{A}_{\ell_f,L}=\int_0^\infty{\rm d}r\,r^3a_{L,0}^{(r),q}(r)R_f(r)R_i(r).$$ The final orbital angular momentum quantum number $\ell_f$ fulfills the condition: $\ell_f=L$. As a consequence, the leading multipole coefficient $a_{L,0}^{q}$ characterizes the atomic transition. Considering the result presented in Fig. 4c), the electric fields of the lower HHs (harmonic order $n<15$) would initiate an even multipole transition ($n_is\rightarrow n_fs$ or $n_is\rightarrow n_fd$). Choosing instead higher harmonics ($n>20$), would result in a dipole transitions, i.e., $n_is\rightarrow n_fp$.}

\section{Conclusions}

HHG  by  focused radially polarized vector fields is dominated by  the  interplay  between the   longitudinal and the transversal  laser components which oscillate with a  $\pi/2$  phase difference and  amplitudes that depend on focusing. Extreme  focusing poses a challenge to experiments as the gradient of the longitudinal component along the optical axis becomes steeper. However the predicted effects are strongest when the longitudinal and transversal components are of comparable strengths. In addition, when averaging over the atoms distributions the atoms right on the optical axis have smaller weight.	Harmonics akin to RVB exhibit  a local circular polarization  \emph{perpendicular} to the focal plane, meaning  that   circular polarized HHGs are producible and tunable  by  varying the beam waist. The discussed  circular polarization relates to the  transverse spin angular momentum’, discussed Ref.[\onlinecite{natphot15}]. Furthermore, multipolar, even and odd  harmonics are  generated. The predicted effects highlight the potential of using structured laser pulses with inherent longitudinal field components for non-linear processes in matter.

\appendix*
\section{Description via LG modes}

\begin{figure}[t]
	\includegraphics[width=0.96\columnwidth]{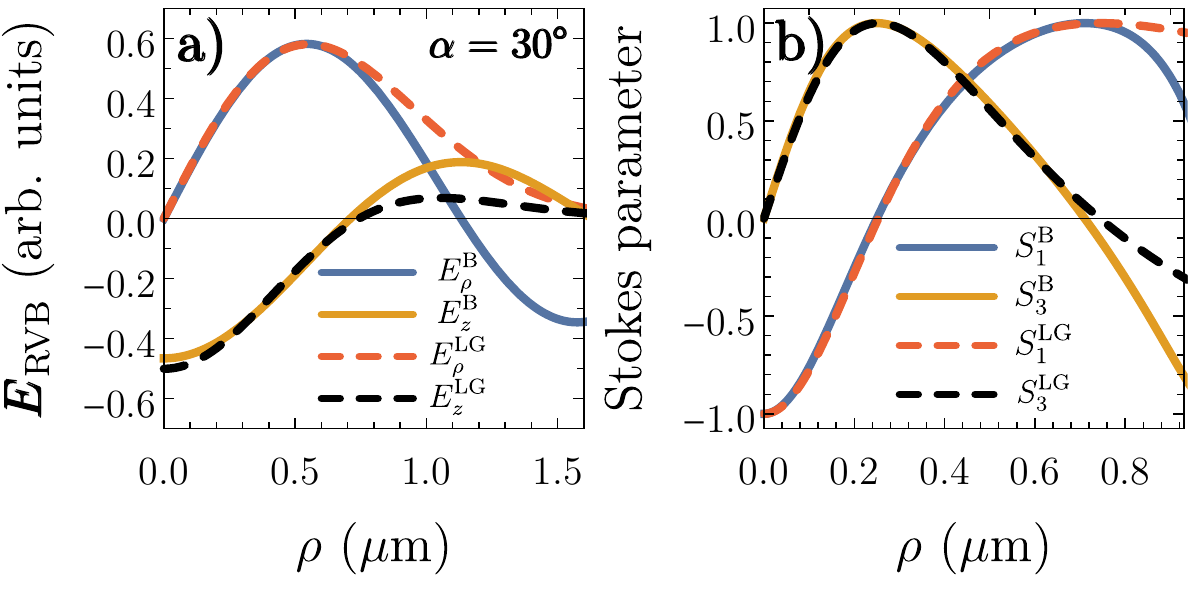}
	\caption{Comparison between RVBs constructed out of Bessel (B) modes and LG modes. a) Transversal and longitudinal fields in dependence on the axial distance $\rho$. b) Spatially-dependent (normalized) Stokes parameters $S_1$ and $S_3$. The opening angle of $\alpha=30^\circ$ belongs to a LG waist size of $w_0=0.72$\,$\mu$m.}
	\label{fig:appendix}
\end{figure}
The vector potential of a Laguerre Gaussian (LG) mode with $|m|=1$ and $p=0$ is given in cylindrical coordinates $\pmb{r}=\{\rho,\varphi,z\}$ by \cite{Allen1992orbital}
\begin{equation}
\begin{split}
\pmb{A}^{m,\sigma}_{\rm LG}(\pmb{r},t)=&\hat{e}_\sigma A_0\mathcal{N}_{\rm LG}\frac{w_0}{w(z)}\frac{\sqrt{2}\rho}{w(z)}e^{-\frac{\rho^2}{w^2(z)}}\\
&\times e^{i\frac{q\rho^2}{2R(z)}+im\varphi+i\zeta(z)}e^{i(qz-\omega t)}+{\rm c.c.}
\end{split}
\end{equation}
Here, $\sigma$ indicates the polarization state with the corresponding vector $\hat{e}_\sigma=e^{i\sigma\varphi}(1,i\sigma,0)^T/\sqrt{2}$. The beam width is $w(z) = w_0\sqrt{1 + (z/zR)^2}$, where $w_0$ is the beam waist while $z_R =qw_0^2$ is the Rayleigh length for a wave number $q=\omega/c$. The function $R(z)$ is the wavefront radius of curvature, given by $R(z) = z[1 + (z_R/z)^2]$ and $\zeta(z)=-2\tan^{-1}(z/z_R)$ is the Gouy phase. The normalization constant $\mathcal{N}_{\rm LG}=\sqrt{2e}$ was chosen in a way that $\pmb{A}^{m,\sigma}_{\rm LG}(\rho_{\rm max},z,t)=A_0$ where at $\rho_{\rm max}$ the peak amplitude can be found.\\
The vector potential of a radial vector beam (RVB) can be found by a sum of $\pmb{A}^{+1,-1}_{\rm LG}(\pmb{r},t)$ and $\pmb{A}^{-1,+1}_{\rm LG}(\pmb{r},t)$ yielding
\begin{equation}
\begin{split}
\pmb{A}^{\rm RVB}_{\rm LG}(\pmb{r},t)=&A_0\mathcal{N}_{\rm LG}\frac{w_0}{w(z)}\frac{\rho}{w(z)}e^{-\frac{\rho^2}{w^2(z)}}\\
&\times e^{i\frac{q\rho^2}{2R(z)}+i\zeta(z)}e^{i(qz-\omega t)}\hat{e}_\rho+{\rm c.c.}
\end{split}
\label{eq:LG_RVB}
\end{equation}
The vector potential is not solenoidal, i.e. via Lorenz gauge condition $\pmb{\nabla}\cdot\pmb{A}^{\rm RVB}_{\rm LG}+\partial_t\Phi^{\rm RVB}_{\rm LG}/c^2=0$ it gives rise to a electromagnetic scalar potential:
\begin{equation}
\begin{split}
\Phi^{\rm RVB}_{\rm LG}(\pmb{r},t)=&A_0\mathcal{N}_{\rm LG}\frac{\omega}{q^2}\frac{w_0}{R(z)w^4(z)}e^{-\frac{\rho^2}{w^2(z)}}e^{i\frac{q\rho^2}{2R(z)}+i\zeta(z)} \\
&\times \left[q\rho^2w^2(z)+2iR(z)(\rho^2 - w^2(z))\right]\\
& \times e^{i(qz-\omega t)} + {\rm c.c.}
\end{split}
\end{equation}
Finally, the associated electric field can be found by $\pmb{E}^{\rm RVB}_{\rm LG}(\pmb{r},t)=-\partial_t\pmb{A}^{\rm RVB}_{\rm LG}(\pmb{r},t) - \pmb{\nabla}\Phi^{\rm RVB}_{\rm LG}(\pmb{r},t)$ resulting in both a transversal $\emph{and}$ longitudinal field component. The electric field in the plane $z=0$ reads explicitly
\begin{equation}
\begin{split}
E^{\rm RVB}_{\rm LG,\rho}&=E_0\mathcal{N}_{\rm LG}\frac{\rho}{q^2w_0^5}e^{-\frac{\rho^2}{w_0^2}} \left[4\rho^2-8w_0^2+q^2w_0^4\right]\sin(\omega t), \\
E^{\rm RVB}_{\rm LG,\varphi}&=0, \\
E^{\rm RVB}_{\rm LG,z}&=2E_0\mathcal{N}_{\rm LG}\frac{1}{q^3w_0^7}e^{-\frac{\rho^2}{w_0^2}} \cos(\omega t) \\
&\quad\times\left[2\rho^4+4w_0^4-q^2w_0^6+\rho^2w_0^2(q^2w_0^2-8)\right].
\end{split}
\end{equation}
where $E_0=A_0\omega$. \\
In Fig.\,\ref{fig:appendix} we show a comparison between the RVB electric fields contructed out of Bessel modes (a) and LG modes (b) in a focused condition, i.e., $\alpha=30^\circ$ which corresponds to a LG waist $w_0=0.72$\,$\mu$m. As presented in Fig.\,\ref{fig:appendix}a), the longitudinal and transversal components show similar trends. While near the optical axis the agreement is remarkable, larger deviations occur behind the first intensity maxima. The reason is the exponentially decreasing field amplitude of the LG mode, while the Bessel beam exhibits infinity side maxima.\\
Important are the Stokes parameters in the $\rho-z$ plane of the driving field, shown in Fig.\,\ref{fig:appendix}b). Here, LG and Bessel RVBs show a remarkable agreement: The zone around the optical axis is ($z$-)linearly polarized (Stokes parameter $S_1\cong-1$), while increasing the axial distance yields a region with a pronounced circular polarization $S_3\cong+1$). Increasing $\rho$ further, results in in-plane polarization (Stokes parameter $S_1\cong+1$), which means the beam is radially polarized in this region. Similar to Fig.\,\ref{fig:appendix}a), LG and Bessel Stokes parameters start to deviate for $\rho>0.8$\,$\mu$m. Since we consider a small interaction volume with an effective radius of 1\,$\mu$m, we expect similar results as reported in Figs. \ref{fig2}-\ref{fig4} when using LG modes instead of Bessel modes for the construction of the RVB.

\begin{acknowledgments}
This research is funded by the Deutsche Forschungsgemeinschaft (DFG) under SFB TRR227, SPP1840 and WA 4352/2-1.
\end{acknowledgments}


%

\end{document}